# Augmenting Actual Life Through MUVEs

Laura Anna Ripamonti, Ines Di Loreto, Dario Maggiorini

D I C O – D e p t .  o f  I n f o r m a t i c s  a n d  C o m m u n i c a t i o n

U n i v e r s i t à  d e g l i  S t u d i  d i  M i l a n o

V i a  C o m e l i c o ,  3 9

I - 2 0 1 3 5  M i l a n o

I t a l y

[ripamonti, ines.diloreto, dario]@dico.unimi.it

# AUGMENTIG ACTUAL LIFE THROUGH MUVEs

**ABSTRACT**

*The necessity of supporting more and more social interaction (and not only the mere information sharing) in online environments is the disruptive force upon which phenomena ascribed to the Web2.0 paradigm continuously bud. People interacting in online socio-technical environments mould technology on their needs, seamlessly integrating it into their everyday life. MUVEs (Multi User Virtual Environments) are no exception and, in several cases, represent the new frontier in this field. In this work we analyze if and how MUVEs can be considered a mean for augmenting communities' – and more in general people's – life. We trace a framework of analysis based on four main observations, and through these lenses we look at Second Life and at several projects we are currently developing in that synthetic world.*

# THE INTERPLAY BETWEEN VIRTUAL AND ACTUAL: IDENTITY, RELATIONSHIP AND PLACE

The relationship between online and offline life (but we rather use *synthetic* and *actual* – see Castranova, 2005 and De Cindio et al., 2008) has been widely studied in recent years, adopting several different approaches and through the lenses of different disciplines (e.g. psychology, computer science, sociology, economy, architecture, etc.).

An exhaustive analysis of each of the aforementioned research branches is almost impossible; nevertheless, within each of them, some key features naturally emerge denoting particular or remarkable facets of the complex relation which binds together the *synthetic* and the *actual* worlds.

Three key concepts, in particular, seem to be fundamental for investigating how synthetic and actual worlds overlap, intersect and interact to "augment" each other, instead of being counterpoised (Mitchell, 2003; Wellman & Haythornthwaite, 2002). These concepts are *identity*, *relationship* and *place*. It is through these dimensions that we analyze how the MUVEs (Multi Users Virtual Environments) – among which synthetic worlds are one of the more "extreme" products of the cyberculture movement – are becoming more and more an extension of people everyday life. MUVEs does not provide their users with an alternate reality, but augment and add "value" (which should be implicit in the notion of augmentation) to their actual life.

Our framework of analysis is based on four major observations:

*Observation 1:* online identity is an extension of personal actual identity, which is socio-culturally constructed and evolves over time in both worlds.

*Observation 2:* online social networks emerge, in the space of possibilities created by the Internet, as extensions of actual ones; in this process "online identities" can be involved as well.

*Observation 3:* synthetic places are the extension of actual, public and private spaces. They augment people's possibility to interact in online social networks and, at the same time, are affected and shaped by social interactions.

*Observation 4:* online identity, relations and places can interact to augment effectively people actual social life. A careful and exhaustive design of the online social environment is required for this to happen: this means that critical factors affecting social interactions among users must be taken very seriously, and need a consistent amount of study, to guarantee the success of a synthetic world.

## Observation 1: Online Identity Is an Extension of Personal Actual Identity

The Cyberculture movement (Markham, 1998; McKenna & Bargh, 1998) assumed that technology allows people to detach from the actual world, inventing a completely different "virtual" identity. This new identity is completely unconnected to the actual one, since the physical/actual world is cast aside when entering the cyberspace. However, it has emerged (see, for instance, Graham, 2002) that personal identity is based on the interaction between physical and virtual elements even when identity is considered in terms of the online world, thus leading to a completely different conclusion compared to the Cyberculture perspective. Indeed, in the actual world, our body is a *mediator* in creating our personal identity, but when the body is abandoned – precisely as in online social interactions – "technology" replaces it. Paraphrasing Marshall McLuhan (1964), we can consider "technology as an extension of man" (Lister et al., 2003). Just as our corporeal bodies are integral to our personal and social lives, digital self-representations (e.g. avatars) are central to our experience in synthetic environments (Polsky, 2001).

In this vein, Manuel Castells says that people with online identities are nevertheless "bound by the desires, pain and mortality of their physical life" (Castells, 2002, 118), while several case studies support the assertion that online identity extends offline identity: see, for instance, the analysis of RumCom.local newsgroups (Rutter & Smith 1999). Hence we can say that identity is socio-culturally constructed for both the virtual and the actual environments.

Identity in the actual world is continuously evolving due to interaction with the multiple socio-cultural contexts we come across during our lives (Maffesoli, 1996). Online, this phenomenon is enforced by the fact that the Internet is intrinsically "global," thus supporting and multiplying worldwide cross-cultural social interactions. However, people's virtual personality tends to stay increasingly the same, or at least to change over time at the same pace as actual personality (Schiano & White, 1998; Becker & Mark, 2002; Cheng et al., 2002). Online and offline impression management works in very similar ways too. The "cyberselves" are built through *presentation*, *negotiation*, and *signification* (Waskul & Douglass, 1997) and evolve over time due to the ongoing interactions with others, exactly like our "actual selves". Studies in this area seem to indicate that, although people like to indulge in some experimentation with their self-projection, identity play decreases with time. In other words, the longer people use online environments, such as e.g. MOOs or chats, the more likely they are to produce self-presentations that are more "authentic" and, even when some "false" element is present in people first online self-presentations, over time "their true self will seep through" (Leary, 1993; Turkle, 1995; Curtis, 1997; Roberts et al., 1996).

**Observation 2: Online Social Network Emerge As Extensions of Actual Ones**

The online world also has relevant effects on *relationships*, on the natural tendency of people to gather in associations, and – more in general – on community life.

These effects can be seen through dystopian or utopian lenses. On the one hand, the Internet is seen as a means to increase social alienation and the erosion of community life (see e.g., Dreyfus, 2001; Putnam, 1995), even though it acknowledgedly helps building social relations, because such relationships cannot be compared to those of actual life, from which they subtract time. On the other hand, the Internet is seen as a social glue binding collective intelligence, the matrix on which the global village germinates and develops (de Kerckhove, 1997).

Both positions appear too deterministic. As is often the case, the truth may lay in the middle: the Internet could be looked upon as a "space of possibilities" supported by technologies that are unable – on their own – to built or disrupt social networks (Wellman, 2005). This happens, as an example, in synthetics worlds, that – by an active exploitation of all our senses – can create a psychological sense of presence, or, in other worlds, the illusion that "I'm *in* the virtual world and not in my house" and, as a consequence, that "I'm *there with other people*" (Biocca, 1997).

**Observation 3: Online Places Are the Extensions of Actual Public and Private Places**

The Internet has tickled the interest of a large number of different disciplines (geography, architecture, urban planning, computer science, etc.), from which alluring suggestions can be drawn about the role of actual vs. synthetic *space* and *place*. The "sense of place" is defined by cultural geographers, anthropologists, sociologists and urban planners as those characteristics that make a place special or unique, as well as those that foster a sense of authentic human attachment and belonging (see e.g. Relph, 1976; Norberg-Schulz, 1980): a well-known phenomenon in human society, in which people strongly identify with a particular geographical area or location. The term space, on the contrary, can be viewed as a set of dimensions in which objects are separated and located, have size and shape, and through which they can move.

In the virtual world, people generate a "sense of place" – exactly as it happens in the actual world (Mitchell, 1995) – and tend to interact with virtual space using the same metaphors adopted for the actual world. Cyberspace, like its actual counterpart, can be zoned, trespassed upon, interfered with, and split up into small landholdings similar to actual property holdings. These effects are sometimes emphasized when they involve online communities: just as actual communities need an appropriate mix of private and public places to prosper, their online versions need analogous places, carefully designed to effectively support the social interactions that underlie community life. It is through the balance of these two types of place that we encourage spontaneous conversation and social-network building among 'neighbors.' Such interaction is the terrain upon which strong relationships, sense of community and identification germinate (Wenger et al., 2002).

**Observation 4: Online Identity, Relations and Places Can Interact to Augment Actual People Social Life.**

We observe that the three concepts – identity, relationship and place – are strongly linked and enforce each other in both environments (the actual and the virtual – see Fig. 1). The use of effectively designed spaces enforces (and is enforced by) the building of social networks – that is to say a net of relationships – but social networks constitute an ideal environment for expressing and evolving personal identities. Last but not least, spaces are shaped by identities and social networks. This implies that an appropriate 'use' and mix of these three elements may serve as a fulcrum to achieve noteworthy results when dealing with online communities.

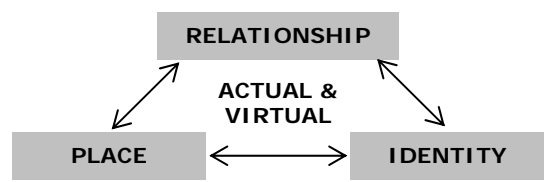

**Figure 1 -** Actual and virtual place, identity and relationship

In the following paragraphs we will see to which extent this can be empirically proved, investigating through these lenses the case of a MUVE: Second Life.

# MUVES, MMORPGS, MUDS AND OTHER SYNTHETIC WORLDS

MUVEs are online, multi-user virtual environments, also known as "virtual worlds". This term has been used till recently to refer to the evolution of more traditional 3D chats, Multi-User Dungeons/Domains/Dimensions (MUDs), MUDs Object Oriented (MOOs) and Massively Multiplayer Online Role-Playing Games (MMORPGs), but now it is widely used to indicate Massively Multiplayer Online Games (MMOGs) that not necessarily present the whole set of characteristics of a game (e.g. they have no specific goals to reach). Among the more known MUVEs we can list Second Life, There and Active Worlds, beside them there are also some intriguing research projects going on, such as Harvard's Rivercity Project (a MUVE aimed at "learning scientific inquiry and $21^{st}$ century skills") and Croquet .

MUVEs derived from the combination of two technologies: virtual reality and text-based chat environments. Traditionally, MUDs were designed for adventure games played by distributed users. Social use of MUDs subsequently developed and, at times, they become environments for chatting. They were commonly referred as *virtual worlds*, whereas, because of the unfortunate history of the "virtual reality" scientific research paradigm, the "virtual" tag was opposed to the "real" one. This is the main reason for which we prefer to call them "synthetic worlds": it conveys an idea not of a different and detached "other" reality (often also connoted with negative meanings), but of something perhaps unnatural, but nonetheless bounded to our everyday life.

Technically speaking, MUVEs are online persistent virtual worlds represented using 3D isometric/third-person graphics, that allow for a large number of simultaneous remote users to interact. This means that they generally offer (more or less) realistic 3D graphics and physics to bring the users in a space populated by objects that may or may not recall those of the actual world. They are not necessarily games, but they are always social environments, inhabited by avatars (usually two or three-dimensional graphical representations of humanoids), that may have "demi-god" abilities, such as being able to fly and change their appearance at will.

If we dig a little deeper in MUVEs characteristics, we discover that they are not simply the last ring of the online social environment evolutionary chain. Some MUVEs (e.g., Second Life) have some evident – and some more implicit – characteristics that subtly trace a fracture between them and the previous generation of MUDs, MMORPGs, etc., and pave the way to an unforeseen possible convergence with the Internet (and the web) communicative potentialities, since they effectively couple content diffusion and social interaction.

## The Synthetic World of Second Life

Second Life (SL for short) is one among several virtual worlds that have been inspired by the cyberpunk literary movement and in particular by Neal Stephenson's novel "Snow Crash" (Stephenson, 1992). SL adopted Stephenson's idea of *Metaverse*, a user-defined world in which people can interact, play, do business, and otherwise communicate. Actually, SL was intentionally designed to be an environment completely constructed by its users (Ondrejka, 2004). Created in 2001 by Linden Lab and launched in the public in 2003, it registered a skyrocketing diffusion, and in a very short period its users outnumbered those of any other similar environment (at the moment SL counts about 7 millions registered users from all over the world, among them more than half a million are very active).

SL users are represented by motional avatars, which are the medium used to interact, explore, socialize, participate in individual and group activities, etc. SL users define themselves as "residents": it is noteworthy that this term suggests an idea of "citizenship". As a matter of fact, early residents strongly felt their belonging to the synthetic world, and they organized in-world[1] public demonstrations to counteract specific policies or rules adopted by Linden Lab they did not agree upon (this happened, e.g., when residents were being charged for objects they created in-world: a protest has been set in-world, sending out a Thoreau-style proclamation against Linden Labs, see Rymaszewski et al., 2007 p. 282).

Since SL was conceived as an empty world, its internal building system is powerful and easy to use (compared to other similar 3D development tools). It allows manipulation of geometric primitives: residents – alone or collaboratively – can mould these "prims" into new shapes, change their texture and physical qualities, link them together for creating objects as complex as they like, add contents (e.g. text, multimedia, etc.) or make them interactive through a scripting language. Content creation in SL involves skills like graphic design, three-dimensional modelling and programming. The ability of users to learn the relatively easily programming language and to create objects on their own made *Second Life* particularly popular. Creation and crafting is an intriguing component of SL: it attracts so many users and has played a relevant role in SL success. Actually, it was by engaging its users in the act of creation that SL produced an environment different from others virtual words: residents become a sort of producer-consumer (similar to the thousands of people who are mixing their own music, making their own movies or publishing their own art or texts on the Internet). Many MUDs and MMORPGs have contents that were – and continue to be – built primarily by their users (Lastowka & Hunter, 2004; Turkle, 1995), but they imply at least two major constraints to creativity: objects and contents should often be tuned with the environment (e.g. medieval or science fiction) and the creator does not have any intellectual property right on them. On the contrary, following a farseeing suggestion by Lessig (Rymaszewski, 2007; Lessig, 2004; Lessig, 2001), SL residents preserve their intellectual property rights on each object or content they create in-world, and these objects can be sold or bought using a synthetic currency (Linden Dollar), that can be traded for US Dollars according to a fluctuating rate of exchange.

**Some Technical Insights About Second Life**

SL is implemented as a client-server system; the clients will connect to a server holding the metaverse content. Content inside the metaverse is made up of basic shapes (named also primitives or prims) which can be linked together to create complex objects. Due to the limited number of prims and the relative ease to describe them, SL can use a relatively low-bandwidth streaming-like system to push environment data first and multimedia (like textures and sounds) later to the client. This system has been proved to improve user experience while being less demanding in term of network resources.

The SL metaverse is not located on a single server, but resides on the implementation of a huge distributed system. This distributed system, or *grid*, is made up from a federation of nodes, each one taking care to simulate the environment inside a given virtual space of 256 by 256 meters. Due to this function, these nodes take the name of "Simulators" (or SIMs for short). Each SIM acts as a virtual machine: it takes in input actions from avatars and objects within, applies them to the current virtual environment together with physical rules and local policies, and provides back a new environment. The global SL grid is the result of a 2D distribution of SIMs, glued together by a global messaging system, where avatars can walk or fly between virtually adjacent SIMs. As for January 2008 the current "geographical" extension of the SL grid is about 26 millions acres.

Despite the fact SL is a completely distributed system, in order to ensure data consistency between SIMs and some other features unique to SL, there are a number of operations performed in a centralized way. Authentication, profile management and economic transactions are managed by a back-end service, whereas in-world objects management is achieved by means of a dedicated server (*asset server*). The asset server is in charge to assign a unique ID to all objects present in-world, to provide consistency for objects uniqueness between SIMs, and to apply access policies to preserve resident's intellectual properties.

---
[1] The word "in-world" is commonly used among SL residents for indicating events taking place into the synthetic world, counterposing to events taking place in RL (real life).

As already mentioned before, in-world objects can be "augmented" by user-created programs using a special purpose programming language (the Linden Scripting Language or LSL). Using LSL, a resident can describe objects reactions to stimuli or interactions, from an avatar or the surrounding environment.

Interaction between avatars and objects is governed by a messaging system, which can be local to a SIM or global to the grid. Local interaction is initiated by the interface (like mouse clicks or keyboard press) and by text messages (like chat). When one of these events is triggered, the simulator will distribute a number of messages to involved avatars and objects; reception of these messages might imply the visualization of a text message and/or a state change for a program inside an object. Global interaction is essentially text-based and is mainly intended as an inter-SIM instant messaging system; both avatar and objects can benefit from global communications.

Communication is a key point of SL: the messaging system can be intertwined with all other in-world operations. Relationship between avatars will extend in-world with no distance boundaries and will also span off to real life, because messages will be relayed via e-mail when the user is not online. Expressing personal identity can be performed not just by avatar reshaping, but also by wearing (attaching) scripted prims, which in turn will be able to interact and send messages to nearby objects and avatars. In this way attachments will be playing a role in how the surrounding environment will perceive the user presence, even at a distance. Places can be filled with interacting and active scripted objects, which will send messages to users no matter where they are, thus, helping creating a social network without the constraint of being "there".

To some extents, SL communication is not limited to the grid itself: scripted objects have means to reach the Internet and use data from it to augment the virtual environment; it is possible to access web content as well as send (and receive) e-mail messages.

Multimedia content from the Internet is supported in an indirect way; a real-time media stream can be set as part of the environment and the SL client will take care to independently retrieve the content and perform the playout without interfering with the grid.

**The It.net and Others Ongoing Projects**

Our experiences with SL has begun during late winter 2006 and firstly concretized in a cycle of seminars in the framework of a course on online communities building for undergraduate students in Computer Science. From this first positive experience (students were enthusiast of having classes and meeting teachers in the synthetic world) bud several projects, among which the more relevant are:

- a project with a local body, aimed at building an in-world presence supporting the activities of its sector devoted to touristic promotion. This project unfortunately aborted few weeks before its official inauguration, due to political discussion about online presence that, in the meantime, grew among public officers;
- a study about how SL can be exploited to sustain and improve companies communication activities. This project is under development in partnership with a company working in the advertisement and marketing industry;
- a project aimed at investigating if and how SL could be a mean for supporting emerging young musicians (thus also comparing it with other very popular tools for

- music sharing). This project involves a not for profit company, whose main goal is helping young musicians in the actual world.
- the It.net project: an ambitious initiative, whose main goal is determining commonalities and differences between approaches necessary at building web-based and MUVE-based communities;
- a collaboration with a course of virtual reality for undergraduate students in Computer Science. As part of their homework, students of this course should create – using advanced 3D graphical editors such as Maya – buildings for the It.net project.

The It.net project was born during late summer 2007 as a comprehensive environment in SL aimed at collecting several in-world experiments under development by a group of students graduating in Computer Science. It consists of an area where different aspects of the synthetic world are explored through different lenses, nonetheless they are knit firmly together by a common idea: the creation of a shared social network.

In the following we will use our experiences in the It.net project (although it is currently still under development) to test our approach to the adoption of SL and – more in general – of specific MUVEs as means for effectively augmenting people's everyday life.

**SECOND LIFE DISTINGUISHING CHARACTERISTICS**

According to our observations, three major aspects are the basis upon which communities can eventually germinate: how people presents their identities, how those identities are used to interact in social networks and, to which degree people and their networks are able to mould spaces into places and are – vice-versa – influenced by them. SL features strongly supports each of these aspects.

**Identity Creation and Management in SL**

MUVEs (and *Second Life* between them) generally effectively support creation and management of online identities. Participants are usually registered members, identified in the synthetic environment by their pseudonym (nickname, username) and by an avatar. Other information about them (such as age, interests, etc.) may be provided and made publicly available in a user profile. The username they choose, the details they do or don't indicate about themselves, the presented information, and the avatar they assume in the online community — all are important clues about how people manage identity in synthetic environments.

Since SL offers a visual environment where practically each detail is customizable, at least two aspects are fundamental for identity building and management: your name and your avatar. In our actual lives these characteristics are persistent (except for very particular cases) and we should adapt to whichever choice *others* have done for us. This is not true for SL: its residents can invent the name that better suits their personalities and adopt or even *create* no matter which self-representation they like. They can be a dog, an elf, a human, a dragon, a can of Coke, etc.; no constraint limits their creativity. Their appearance can be further customized by adding e.g. special textures, clothing, and "animation overriders" (scripts that add much more natural movements to the avatars). These features of SL are very relevant and "make the difference" with other MUVEs and MMOGs – where users can only *choose* their identity among a set of pre-defined avatars and change their clothing – since, like in the actual world, the avatar/body is the "suit" used for self-presentation in social environment.

Avatars choices in SL, however, generally conform to cultural standards of what is considered attractive or normative (Lastowka & Hunter, 2004): that is to say, the particular cultural view of the more influential or numerous groups of users impacts the virtual space. This is largely visible in SL*,* where is quite difficult to find an avatar that really diverges from the standards.

As a matter of fact, it is not straightforward to undertake a conversation with a puzzling avatar such as a flying metal ball, while a plump inoffensive teddy is by far more reassuring.

While SL residents can change their appearance as many times as they like, they are not allowed to change their avatar's name: a name chosen at registration time is *the* name, and the only way to reappear in SL with a new identity is creating a new account. SL identities should be composed by a name *and* a surname: while residents can pick up whichever name they fancy about, they have to select their surname from a (very long) list provided by Linden Labs. This procedure have some implications: the similarity to what happens in actual life (at least in western cultures) makes the choice of names such as e.g. "amy48" or "starry_night_47" – a normal praxis for email addresses – sound quite "strange" in SL; moreover, aroused the necessity to bring, under some circumstances, actual names into *Second Life*. It's the case of the novelists Ellen Ullman and Cory Doctorow, of the game designer Harvey Smith, creator of the games Deus Ex; again, of the singers Suzanne Vega and Duran Duran (who appeared as themselves in an island on which they performed live concerts), and the politicians Mark Warner and Hillary Clinton.

A publicly accessible profile is associated to each SL resident. Profiles are a powerful mean of self-presentation and impression management: they are essential to declare to the world who you are, which are your interests, what your avatar looks like, and what you think is worth of seeing in SL. They may even contain details of your actual life (your name, contact details, portrait, etc.). Profiles, last but not least, inform about the groups you belong to and about your favourite places in SL, and may contain advertising about your business and "profession" in SL.

Although explicit tools for supporting reputation building and tracking are not provided by the SL client, it is undeniable that the combination between identity representation (name, avatar customization, user profile information) and the retention of the creator/owner by user created contents create a powerful mix, able to strongly characterize through virtuous circles residents both in the synthetic and in the actual world.

The main features of SL that act as enabler for effective identity creation and management are summarized in Tab.1.

*Table 1. Several SL synthetic world distinguishing characteristics – Identity*

| SL Characteristic | Notes/implications |
| --- | --- |
| *Avatar detailed personalization* | SL residents can deeply *customize* their identity, while users of other MUVEs can only *choose* their identity among a set of pre-defined avatars. |
| *Unconstrained avatar personalization* | SL residents can create and adopt whichever representation they like for their avatars, no constraint exist. |
| *Resident detailed profile* | A publicly accessible profile is associated to each SL avatar. It contains information about the resident that can be automatically generated (e.g. groups subscribed) or provided by the users herself (e.g. information about real life, about favourite sites in SL, etc.). |
| *Persistent user-chosen name and identity* | When a new account is created in SL, the user chooses a name (whichever) and picks up a surname in a pre-defined list of several hundreds. This name cannot be changed for any reason and will be indissolubly linked to the avatar and to every object she eventually creates. |
| *Gestures and animations* | Users-created gestures and animations can be applied to the avatar, further personalizing it. |

**Relationships Creation and Management in SL**

SL is a synthetic society where residents engage in a multiplicity of different activities and are involved in a variety of social relations. Similarly to what happens in the actual society, SL social relations can be of different types: some more formal than others, some transitory and some other connected to friendship networks. For many SL residents the synthetic world is simply a place to hang around and meet new friends, for others is a place for gaming or doing business. Residents organize all sorts of events in SL: movie festivals and shows, scientific conferences, parties, literary meetings, etc. All these activities are supported by an appropriate set of socio-technical features, that impact both at individuals and groups level.

From the individual point of view, the creation and management of social relations is enabled primarily by the elements discussed in the previous paragraph, that are further enforced by other specific features of avatars. As an example, SL avatars can use gestures: a gesture is a 3D implementation of chat emoticons, that is to say a way to support *phatic communication* (Stewart & Williams, 2005; Caron & Caronia, 2001), thus reinforcing linkages among people and building common grounds upon which interaction can take place easier (Rintel & Pittam, 1997; Bickmore & Picard, 2005). Technically, gestures are a combination of animation, pose, text and sound. Once assembled, residents can use a gesture by triggering it via text or shortcut keys. Users-created gestures and animation can be applied to the avatar, further personalizing it, and making it a bit more resembling an actual being. Other relevant affordances for social interaction are more "common" tools such as chats (SL client supports both textual and voice chat), instant messaging, buddy list, online presence indicators, etc.

Social interaction among individuals inevitability leads to the creation of groups and the consequent agreement onto a set of shared behavioural norms. Harrison and Dourish (1996) pointed out that the appropriateness of social behaviour in a particular multi-user virtual environment depends on the interpretation of it by individual participants and on the social construction of knowledge. Similarly to what happens in other online communities, SL has rules and policies that limit residents activities. A fundamental set of formal rules (the so-called "six big no-no") must be signed by every new resident when subscribing to the service. These rules are valid all over the synthetic world, but, beside them, other formal rules – usually defined by users or groups – can regulate behaviour e.g., in specific regions or among specific social networks. This is precisely what we have done for regulating students behaviour during the classes in our earlier experiment: we defined a specific netiquette residents whishing to visit our area (during or outside class hours) are expected to respect. In general we could say that different places in SL are devoted to different activities, supported by different groups and, thus, are regulated by more or less formal rules, that can vary between very simple netiquette (the "six big no-no") to very complex structures (sometimes documented in appropriate libraries and supported by classes in "proper behaviour" – as it is the case of the Mentor group).

Groups are generally created by residents, and collect people sharing similar interests. As actual groups, they are a collection of members playing different roles, and endowed with certain special privileges, including sharing land and money. They can build in the land owned by the group, and communicate in a more private way, using group internal messaging system. Similarly to what happens for individual identity, groups too have profiles, that can be partially or totally public. Group profiles contain information about the group (logo, mission, etc.), members list, shared notices and activities, polling tools, etc. The subscription to the group can be open or restricted and for free or subdued to a fee. Belonging to a group can be explicitly shown: a member title can be made publicly visible near the avatar name. This group visibility impacts also on identity creation and management, as well as land sharing in a group impacts on places management.

It is noteworthy that social interaction taking place in SL is supported also by tools that are not in-world. Many discussions about SL take place in web-based forums, and can include knowledge that exist inside or outside the boundaries of the synthetic world. Residents have also created several tools that – in perfect Web2.0 style (O'Reilly, 2005) – allow to import and export contents from/into SL. A website called SLProfiles acts as a kind of MySpace for SL residents, Snapzilla is the SL version of Flickr, BlogHUD allows SL people to post directly to their blogs, and so on, in a perpetual attempt to create a seamless conjunction between in-world and the rest of the Internet.

The interplay between actual and virtual relationships in SL emerges also from several residents projects, such as the "Better World Island" which touching exhibits about life in a Darfur refugee camp. A number of renown not-for-profit organizations, including Techsoup.org, Creative Commons, and Omidyar Network have their in-world "versions".

The main features of SL that act as enabler for effective relationships creation and management are summarized in Tab. 2.

*Table 2. Several SL synthetic world distinguishing characteristics – relations*

| SL Characteristic | Notes/implications |
|---|---|
| *Support to social networks* | Residents' social network is supported by an effective variety of tools (e.g. friends lists, sharing of objects, groups creation by users, access lists, etc.). |
| *User-defined groups* | Groups can be created and managed by users. No constraints are imposed by the Linden Labs, except that at least two residents subscribe to the group. |
| *Netiquette (customizable)* | Every new resident is required to accept a general netiquette when subscribing to SL. Moreover, specific netiquettes created by residents can apply to specific areas or groups. |
| *Gestures* | Avatars can partially support phatic communication by using gestures. |

**Places Creation and Management in SL**

Form Tab. 3 it is quite immediately perceivable that – from a spatial perspective – actual life and the synthetic world can overlap and interact. Actually, the notions of places and spaces – as conceived by architects and urban planners – can be applied (quite) straight away to the SL environment: anyway, in SL the notion of *space* progressively looses its meaning in favour to the one of *place*. This fact is due to several intertwined features of SL and how they interact and intersect with residents' behaviour. New land in SL is born naked: residents create the whole content of every SL island, and are endowed with the capability to customize even the tiniest details of their land. Residents can design not only buildings, but also oro-geographical and weather conditions of their land, and the flora and fauna it contains. Moreover, avatars do not need a lot of the infrastructures that are indispensable to human beings (such as streets, kitchens, bathrooms, elevators, heating systems, etc.): on the contrary, they can represent an incongruent burden to avatars activities. In other words, practically no constraint limits residents possibility to mould – and re-mould in any moment – synthetic spaces into synthetic places, thus creating environments that intrinsically satisfy not only a quantity of their needs and desires, but also those typical of a community (e.g. the presence of both private and public places – Wenger et al. 2002).

Moulding of spaces into places can take place on different levels: public (any place publicly accessible, such as a square, a library, a town, a garden, etc.), private (a private house – as far as the usual concept of "house" can be applied to SL private spaces) and group. A special attention should be devoted to this latter category, since SL customization tools offers new powerful opportunities to groups. The whole SL is a wonderful and extreme example of Participatory Design (see i.e. Nygaard, 1983; Schuler & Namioka, 1993; Blomberg & Kensing, 1998) and participatory development, that coalesces into projects such as Neualtenburg: an attempt to simulate the look and feel of a functioning Bavarian city. Collaboration and co-design are so stressed in SL, that strong groups tends to create spaces (also public) that deeply mirror the set of values they share. An evident example of such collective sense-making activity and mutual intelligibility is the vast group of islands owned by the Elf Circle, one among the more active and numerous community of SL. Residents belonging to the Elf Circle only occasionally are lovers of role playing games, instead they tend to share similar ethical and cultural values (e.g., respect for nature and other living things, love for literature, poetry, and arts in general, etc.) and this is undeniably reflected by the aspect of their lands, that are dominated by beautiful landscapes and buildings, where no explicit representation of technology is allowed. Thus, only by visiting those areas, "foreign" residents are immediately immersed in a well-defined atmosphere.

Other phenomena well-known to architects are reproduced into SL synthetic spaces. As an example, people tend to redesign or reallocate pre-defined spaces to better fit with their needs. This is precisely what happens, e.g., in the "Help" and "Orientation" islands, the spaces where new-born residents arrive when entering SL for the very first time. A special category of long-date volunteer residents (the Mentors) hang around in those island for lending a hand (and their tacit knowledge – Nonaka & Takeuchi, 1995) to newbies; nonetheless these areas have also become a place where mentors meet and enforce their mutual network relations. In the same vein, the *recombination* of places (see e.g. Aurigi & Graham, 2002) is a very frequent event in SL: shopping area are seamlessly coupled with graceful hamlets, plants of a botanical garden are also for sale, etc.

Last but not least, people using SL often experience a sort of "double belonging" that mixes together the actual and the synthetic places: for example, residents can interact through avatars present in a synthetic places while sitting in an actual place and discussing with other residents about actual life or work life issues, as it is often the case in our work-group at It.net, and as it happened recently, when IBM employees went on strike both in the actual and in the synthetic worlds (IBM owns several islands in SL). This behaviour matches with recent evolution (see Soukup, 2006) in the concept of online third place (see Oldenburg, 1989a; Oldenburg, 1989b; Oldenburg & Brisset; 1982), according to which online third places (i.e. online communities) are sustained by Internet technology in multiple actual places. In this situations people can bring a synthetic third place with them during their everyday life and access it form a multiplicity of actual places (e.g. home, office, third places, etc.).

The main features of SL that act as enabler for effective places creation and management are summarized in Tab. 3.

*Table 3. Several SL synthetic world distinguishing characteristics – spaces and places*

| SL Characteristic | Notes/implications |
| --- | --- |
| *Possibility to change spaces into places* | The transition from spaces to places is easy, thanks to the high customizability of the environment (e.g. creation of mountains and rivers, definition of weather conditions and flora, design of buildings, etc.). |
| *Private and public places* | The access of SL virtual land parcels can be open to public or restricted to specific lists of residents. |
| *World Map* | Both spaces and places (e.g. events, etc.) can be retrieved through a map of SL virtual geography, coupled with search capabilities. |

| *Georeferencing* | Search results can be highlighted on the map. |
|---|---|
| *Unconstrained building* | SL residents can build whatever they like in SL. No constraints (e.g., about the architectural style) exist. |
| *Multimedia contents linked to lands* | Multimedia contents (e.g. music or movies) can also be defined by the users as a characteristics of a specific area (e.g. when entering a region a certain music is diffused). |
| *Import of off-world contents* | As it happens in the Web 2.0 paradigm, specific tools can be created to import/export and distribute contents from/to external applications (e.g. RSS readers, etc.). |
| *Collaborative building* | Possibility to grant or deny modify permissions on own objects and buildings to other residents. |

## CONCLUSIONS AND FUTURE DEVELOPMENTS: IDENTITIES, RELATIONSHIPS AND PLACES IN SECOND LIFE

The It.net project is aimed at building a community; hence it has to approach the interplay among the three fundamental dimensions – identity, relations and places – in a comprehensive and holistic way.

In the project, personal identity is perhaps the less analysed dimension, since it is under the direct control of each single resident and, therefore, it is quite impossible (and useless) trying to affect it. More intriguing hints tickle our interest in the remaining two dimensions, their mutual interactions and their interplay with their actual counterparts. Hence, our efforts have a double focus: building a lively social network and effectively binding it to a place. These goals require to sustain social relations and to design spaces that can be easily moulded into places.

The first issue has been addressed in several ways: in order to favour communication and linkages among people (no community can be created if not based on an existing social network – see Wenger et al., 2002) a group has been created to collect all the residents interested in the project. Several events take place on the group land: live music concerts, literary meetings, charity markets, etc. Events are promoted through different media: the group in-world private messages channel, a group blog, message boards on the group land, etc. in order to reach not only members, but also any resident potentially interested, and to give a certain visibility to the group activities. The identity of the group has been enforced by creating a logo, a netiquette and a motto that match with its mission. Moreover, communities are built also on shared knowledge (both tacit and explicit); hence, they require shared repositories of memories. SL technical infrastructure is quite lacking form this point of view, since it provides no effective tools for keeping tracks of communities history and poor tools for the retrieval of contents. The only available alternative has been the creation of a web-based group blog equipped with RSS feeds.

The development of the social network has been intertwined with the design of the places supporting It.net community activities. The destination of the space has undergone a detailed analysis from different perspectives, since we had to couple technical constraints (many buildings mean too many prims, which in turn create lag[2], resulting in a poor user-experience) with the creation of the "sense of place". It.net land contains areas:

---

[2] The "lag" is one of the more feared problems by SL residents. The term indicates the delay in the environmental rendering when entering too crowded or over-built areas. It depends from a variety of factors, among which network latency and bandwidth usage.

- *at a different level of privacy*: public places (connoted by the group identity), private places (teachers' houses) and semi-private places (meeting rooms accessible only to specific residents);
- *with different purposes*: some places are devoted to community activities (e.g., the amphitheatre), while other are more "institutional" (e.g., the area containing multimedia information about students' projects).

The intermediate results of the It.net project, jointly with the lessons learned from the other SL experiences we are developing, seem to support the intuition that the concept of augmentation encompasses the idea of enhancing actual world by seamlessly adding layers of digitally supported value. This form of augmentation is clearly perceivable in the synthetic world of *Second Life*, where, as outlined before, identity, relationship, and place become natural extensions of the actual world.

Thanks to augmentation, the value perceived by SL residents is increased not only along each dimension, but also by their mutual interplay. In Fig. 2 we sketched the superimposed interactions that take place, on the one hand, among identity, relations and place, and on the other hand between their actual and synthetic expression. Online identity, relation and place extend their offline counterparts; similarly, rules that regulate their interactions behave in the same way for both the synthetic and the virtual worlds. This double circular interaction can support people actions in both worlds in an effective way, by creating a technology-enabled environment, appropriate for augmenting social interaction.

Next steps in our research will focus on analyzing results of the It.net and the other projects under development, with a special attention on the impact of design choices on community development in synthetic world and on the differences existing among 2D (i.e. web-based) and 3D communities. Since technology-enabled communities are socio-technical systems, we will consider the impact of design choices on both aspects.

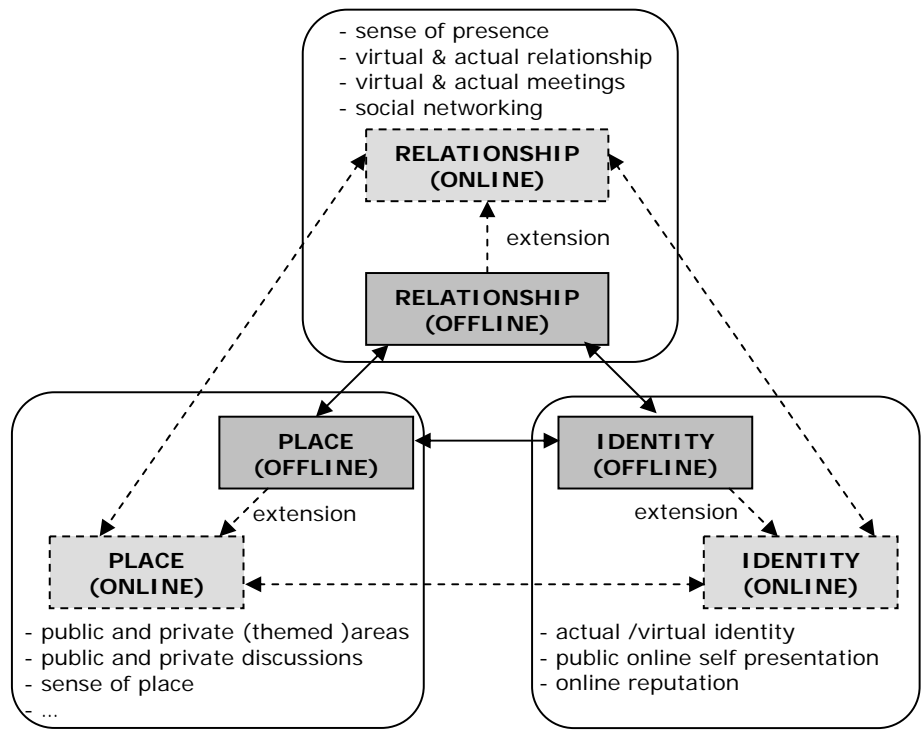

**Figure 2 -** Relation between actual and virtual aspects in Second Life

**Acknowledgements**

We wish to thank Andrea Spirolazzi, Francesca Piazza and Paolo Mariotto. Without their efforts and work our investigations and knowledge of SL would have been by far less effective.

*References*

Aurigi, A., & Graham, S. (2000). Cyberspace and the city: the virtual city in Europe. In G. Bridge and S. Watson (Eds), *A companion to the city*, 489-502. Oxford: Blackwell.

Becker, B., & Mark, G. (1999). Constructing Social Systems through Computer-Mediated Communication. *Virtual Reality, 4,* 60-73.

Bickmore, T. W., & Picard, R. W. (2005). Establishing and maintaining long-term human-computer relationships. *ACM Transactions on Computer-Human Interaction, 12(2),* 293–327.

Biocca, F. (1997). The cyborg's dilemma: Progressive embodiment in virtual environments. *Journal of Computer-Mediated Communication* [On-line serial], 3(2).   Retrieved on January 2008 from: http://jcmc.indiana.edu/vol3/issue2/biocca2.html

Blomberg, J., & Kensing, F. (1998). Participatory Design: Issues and Concerns. *The Journal of Collaborative Computing*, Special Issue on Participatory Design, Computer Supported Cooperative Work, 3-4, 167-185.

Caron, A. H., & Caronia L. (2001). Active users and active objects: The mutual construction of families and communication technologies. *Convergence: The International Journal of Research into New Media,* 7(3), 38-61.

Castells, M. (2002). *The Internet galaxy: Reflections on the Internet, Business, and Society*. Oxford, UK: Oxford University Press.

Castronova, E. (2005). *Synthetic Worlds: The Business and Culture of Online Games*. Chicago, IL: University of Chicago Press.

Cheng L., Farnham S., & Stone L.(2002). Lessons learned: Building and deploying shared virtual environments. In R. Schroeder (Ed.), *The Social Life of Avatars*, 90–111. London: Springer.

Curtis, P. (1997). Mudding: Social phenomena in text -based virtual realities. In S. Kiesler (Ed.), *Culture of the Internet*, 121-142. Hillsdale, N.J.: Erlbaum.

De Cindio, F., Ripamonti, L.A. &  Di Loreto, I. (2008 - forthcoming). The Interplay Between the Actual and the Virtual citizenship in the Milan Community Network Experience. In A. Aurigi and F. De Cindio (Eds.), *Augmented Urban Spaces: Articulating the Physical and Electronic City*. Aldershot, UK: Ashgate.

de Kerckhove, D. (1997). *Connected intelligence: The arrival of the web society*. Toronto, Canada: Somerville House.

Dreyfus, H. L. (2001). *On the Internet*. New York, US: Routledge.

Graham, M. (2002). *Future Active*. New York, US: Routledge.

Harrison, S., &  Dourish, P. (1996). Re-place-ing space: The roles of space and place in collaborative systems. *Proceedings of CSCW 96*, 67-76. New York, NY: ACM.

Lastowka, F.G & Hunter, D. (2004). The Laws of the Virtual Worlds. *California Law Review, 92(1),* 3-73.

Leary, M.R. (1993). The interplay of private self-processes and interpersonal factors in self-presentation. In J. Suls (Ed.), *Psychological perspectives on the self*, 127-55. Nahwah, NJ: Erlbaum.

Lessig, L. (2001). *The Future of Ideas: The Fate of the Commons in a Connected World.* Random House: New York.

Lessig, L. (2004). *Free Culture.* New York: The Penguin Press.

Rymaszewski et al. (2007). *Second Life: The Official Guide*. New Jersey, US: Wiley.

Lister, M., Kelly, K., Dovey, J., Giddings, S., & Grant, I. (2003). *New Media: A Critical Introduction*. London, UK: Routledge.

Maffesoli, M. (1996). *The Time of the Tribes*. London, UK: Sage.

Markham, A. N. (1998). *Life Online: Researching Real Experience in Virtual Space*. Walnut Creek, CA: AltaMira.

McKenna, K. Y. A. & Bargh, J. A. (1998). Coming out in the age of the Internet: Identity "Demarginalization" through virtual group participation.  *Journal of Personality and Social Psychology, 75(3),* 681-694.

McLuhan, M.(1964). *Understanding Media*. New York, US: McGraw-Hill.

Mitchell, W. J. (1995). *City of bits—space, place and the infobahn*. Cambridge, Ma: MIT Press.

Mitchell, W. J. (2003). *Me++: The Cyborg Self and the Networked City*. Cambridge, Ma: MIT Press.

Nonaka, I., & Takeuchi, H. (1995). *The Knowledge-Creating Company.*  New York: Oxford University Press, Inc.

Norberg-Schulz, C. (1980). *Genius Loci: Towards a Phenomenology of Architecture*. New York: Rizzoli.

Nygaard, K. (1983). Participation in System Development. The Task Ahead. In U. Briefs, C. Ciborra, & L. Schneider (Eds.), *Systems design for, with and by the users*. North-Holland.

Oldenburg R. (1989a). *The great good place: Cafes, coffee shops, community centers, beauty parlors, general stores, bars, hangouts, and how they get you through the day*. New York: Paragon House.

Oldenburg, R. (1989b). *Celebrating the third place: Inspiring Stories about the Great Good Places at the hearth of our communities*. New York: Marlowe & Company.

Oldenburg, R., & Brissett, D. (1982). The Third Place. *Qualitative Sociology, 5(4),* 265–284.

Ondrejka, C. (2004). Escaping the Gilded Cage: User Created Content and Building the Metaverse. *New York Law School Law Review, 49(1),* 81-101.

O'Reilly, T. (2005). Web 2.0: Compact Definition?, O'Reilly Radar blog, 1 October 2005. Retrieved on January 2008 from: http://radar.oreilly.com/archives/2005/10/web_20_compact_definition.html.

Polsky, A. (2001). Skins, patches and plug-ins: Becoming women in the new gaming culture. *Genders,* 34(Fall). Retrieved November 28, 2007, http://www.genders.org/g34/g34_polsky.html

Putnam, R. D. (1995). Tuning in, tuning out: The strange disappearances of social capital in America. *PS: Political Science and Politics, 28,* 664-683.

Relph, E. C. (1976). *Place and Placelessness*. London: Pion Publishers.

Rintel, E. S., & Pittam, J. (1997). Strangers in a strange land interaction management on Internet Relay Chat. *Human Communication Research, 23(4),* 507-534.

Roberts, L. D., Smith, L. M., & Pollock, C. (1996). Exploring virtuality: Telepresence in text-based virtual environments. Paper presented at the *Cybermind Conference*, Curtin University of Technology, Perth, Western Australia.

Rutter, J., & Smith, G. (1999). Presenting the Off-line Self in Everyday, Online Environment. *Identities in Action Conference*, Gregynog, UK.

Schiano, D., & White, S. (1998). The first noble truth of CyberSpace: people are people (even when they MOO), Proceedings of the *SIGCHI conference on Human factors in computing systems*, 352-359, Los Angeles, California, United States.

Schuler, D., & Namioka, A. (1993). Participatory Design: Principles and Practices. Hillsdale, NJ: Erlbaum.

Soukup, C. (2006). Computer-mediated communication as a virtual third place: building Oldenburg's great good places on the World Wide Web. *New Media Society 8(3),* 421–440.

Stephenson N. (1992). *Snow Crash*. New York: Bantam Books

Stewart, K., & Williams, M. (2005). Researching online populations: the use of online focus groups for social research. *Qualitative Research, 5(4),* 395-416.

Turkle, S. (1995). *Life on the Screen: Identity in the Age of the Internet.* New York, US: Simon & Schustermany.

Waskul, D. & Douglass, M. (1997). Cyberself: The emergence of self in on line chat. The Information Society, 13 (4), 375-396.

Wellman, B. (2005). Connecting community: On- and offline. Retrieved on January 2007 from: http://www.chass.utoronto.ca/~wellman/publications/index.html

Wellman, B., & Haythornthwaite, C. (2002). *Internet in everyday life*. Oxford: Blackwell.

Wenger, E., McDermott, R., & Snyder, W. M. (2002). *Cultivating communities of practice - A guide to managing knowledge*. Boston, MA: Harvard Business School Press.

**Useful links**

Active Worlds < www.activeworlds.com/>, accessed 31 January 2008

BlogHUD <http://bloghud.com/>, accessed 31 January 2008

Creative Commons <http://creativecommons.org/>, accessed 31 January 2008

Croquet <www.opencroquet.org>, accessed 31 January 2008

Harvard's Rivercity Project < http://muve.gse.harvard.edu/rivercityproject>, accessed 31 January 2008

IT.net Blog < http://itnet-sl.blogspot.com/>, accessed 31 January 2008

Maya <http://usa.autodesk.com>, accessed 31 January 2008

Omidyar Network <http://home.omidyar.net/>, accessed 31 January 2008

Second Life <www.secondlife.com >, accessed 31 January 2008

Second Life Profiles <www.slprofiles.com/>, accessed 31 January 2008

Snapzilla <http://sluniverse.com/pics/>, accessed 31 January 2008

Techsoup <www.techsoup.org/>, accessed 31 January 2008

There <www.there.com/>, accessed 31 January 2008

**Useful SLurls**

Better World <http://slurl.com/secondlife/Better%20World/128/128/0>, accessed 31 January 2008

Creative Commons<http://slurl.com/secondlife/Kula%204/147/87/25>, accessed 31 January 2008

ElvenGlen <http://slurl.com/secondlife/ElvenGlen/128/128/0>, accessed 31 January 2008

IBM <http://slurl.com/secondlife/IBM/128/128/0>, accessed 31 January 2008

ITnet - LIC - D.I.Co. – UniMi <http://slurl.com/secondlife/pesca/87/81/29>, accessed 31 January 2008

Neualtenburg, Altenburg <http://slurl.com/secondlife/Funadama/101/156/33 >, accessed 31 January 2008

Omidyar <http://slurl.com/secondlife/Omidyar/128/128/0>, accessed 31 January 2008

Orientation Island 1 <http://slurl.com/secondlife/Orientation%20Island%201/123/134/26>, accessed 31 January 2008

Orientation Island Public <http://slurl.com/secondlife/Orientation%20Island%20Public/123/134/26>, accessed 31 January 2008

Techsoup.com <http://slurl.com/secondlife/Plush%20Nonprofit%20Commons/86/94/26>, accessed 31 January 2008